\documentclass[twoside,12pt]{article}
\usepackage{graphicx}
\usepackage{bm} % bold math
\usepackage{amssymb,amsmath,cite,color}

\newcommand{\be}{\begin{equation}}
\newcommand{\ee}{\end{equation}}
\newcommand{\bea}{\begin{eqnarray}}
\newcommand{\eea}{\end{eqnarray}}

\newcommand{\fsi}{FSI }
\begin{document}

\title{ \vspace{1cm} Strangeness Production in Hadron Reactions }
\author{H. Machner,$^{1}$ F. Hinterberger,$^2$ R. Siudak$^3$\\
\\
$^1$FB Physik, Univ. Duisburg-Essen, Duisburg, Germany\\
$^2$Helmholtz-Institut f\"{u}r Strahlen- und Kernphysik der Universit\"{a}t
Bonn, \\ Bonn, Germany\\
$^3$ Institute of Nuclear Physics, Polish Academy of Sciences,
Krak\'{o}w, Poland}

\maketitle
\begin{abstract} The paper gives an overview of strangeness-production experiments at the Cooler Synchrotron COSY.
Results on kaon-pair and $\phi$ meson production in
$pp$, $pd$ and $dd$ collisions,
hyperon-production experiments
and $\Lambda p$
final-state interaction studies are presented as well as a search for a strangeness $S=-1$ resonance in the $\Lambda p$ system.
\end{abstract}
%\eject
%\tableofcontents
\section{Introduction}
We will concentrate on experiments performed on the cooler
synchrotron COSY \cite{Mai97} at the Forschungszentrum J\"{u}lich in
Germany. It can accelerate protons and deuterons up to about 3.7
GeV/c thus allowing $K\bar{K}$ production as well as associated
strangeness production in $pp$ interactions. Although polarized
beams are available, we will concentrate here on experiments making
use of unpolarized beams. Excellent beam quality can be achieved
using electron- and/or stochastic cooling. COSY can be used as an
accelerator for external target experiments and as storage ring for
internal target experiments. The strangeness production experiments
have been performed at the internal spectrometer ANKE by the
COSY-ANKE collaboration, at the internal COSY-11 spectrometer by the
COSY-11 collaboration, at the external TOF facility by the COSY-TOF
collaboration and at the external BIG KARL spectrometer by the
COSY-MOMO and COSY-HIRES collaborations.

We will first discuss kaon-pair production followed by $\phi$ production close to threshold. Then hyperon production in $pp$ and $pd$ reactions will follow. Finally $\Lambda p$ interaction and search for a strangeness $S=-1$ dibaryon are discussed.

\section{Kaon Pair Production}\label{sec:Kaon-Pair-Product}
A wealth of data in two-kaon production is measured in $pp$, $pn$, $pd$ and $dd$ reactions. The total cross sections are compiled in Fig. \ref{Fig:Kaon_pair} as function of the excess energy $\epsilon$.
\begin{figure}[h]
\begin{center}
\includegraphics[width=0.4\textwidth,angle=0]{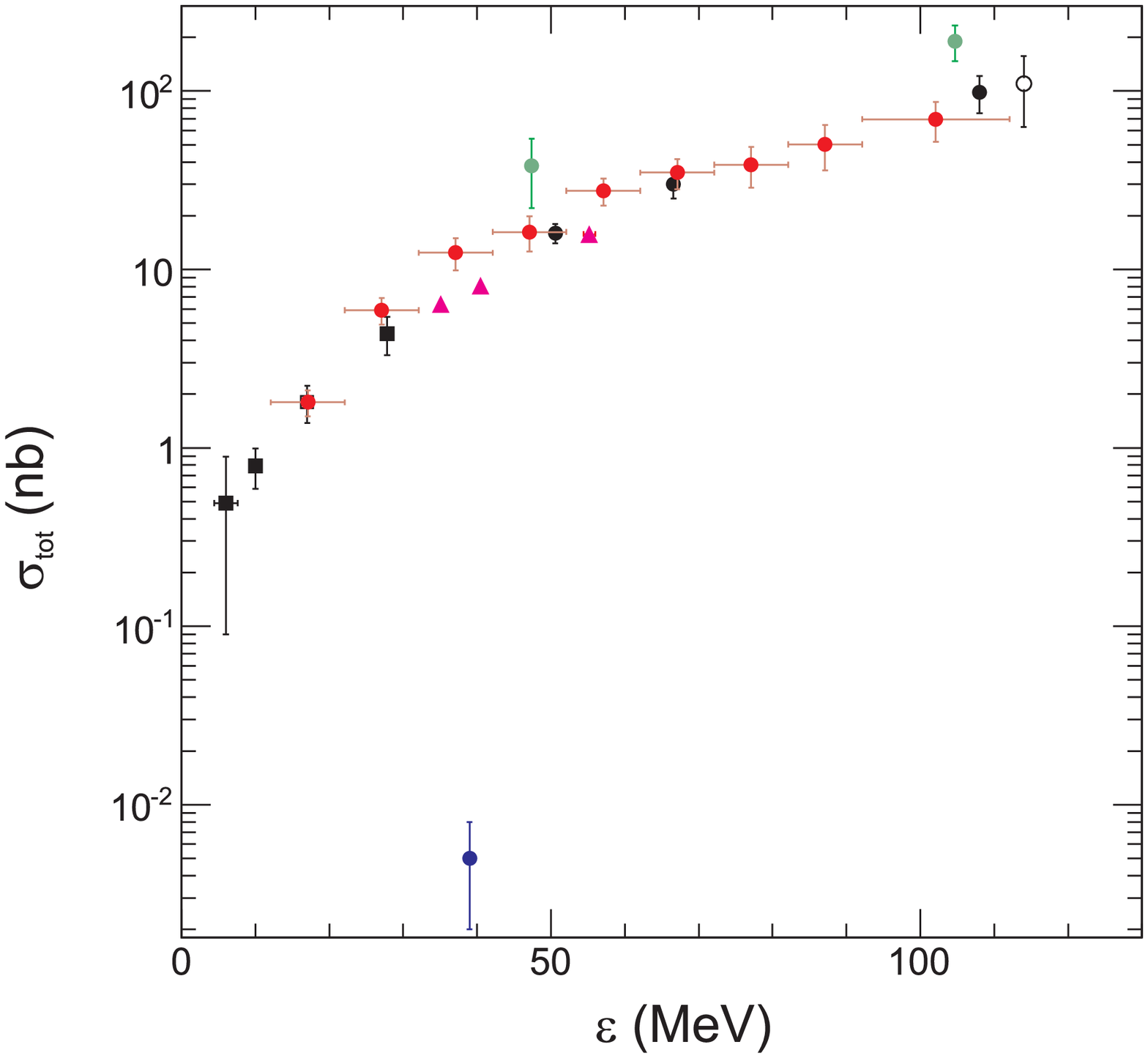}
\includegraphics[width=0.4\textwidth]{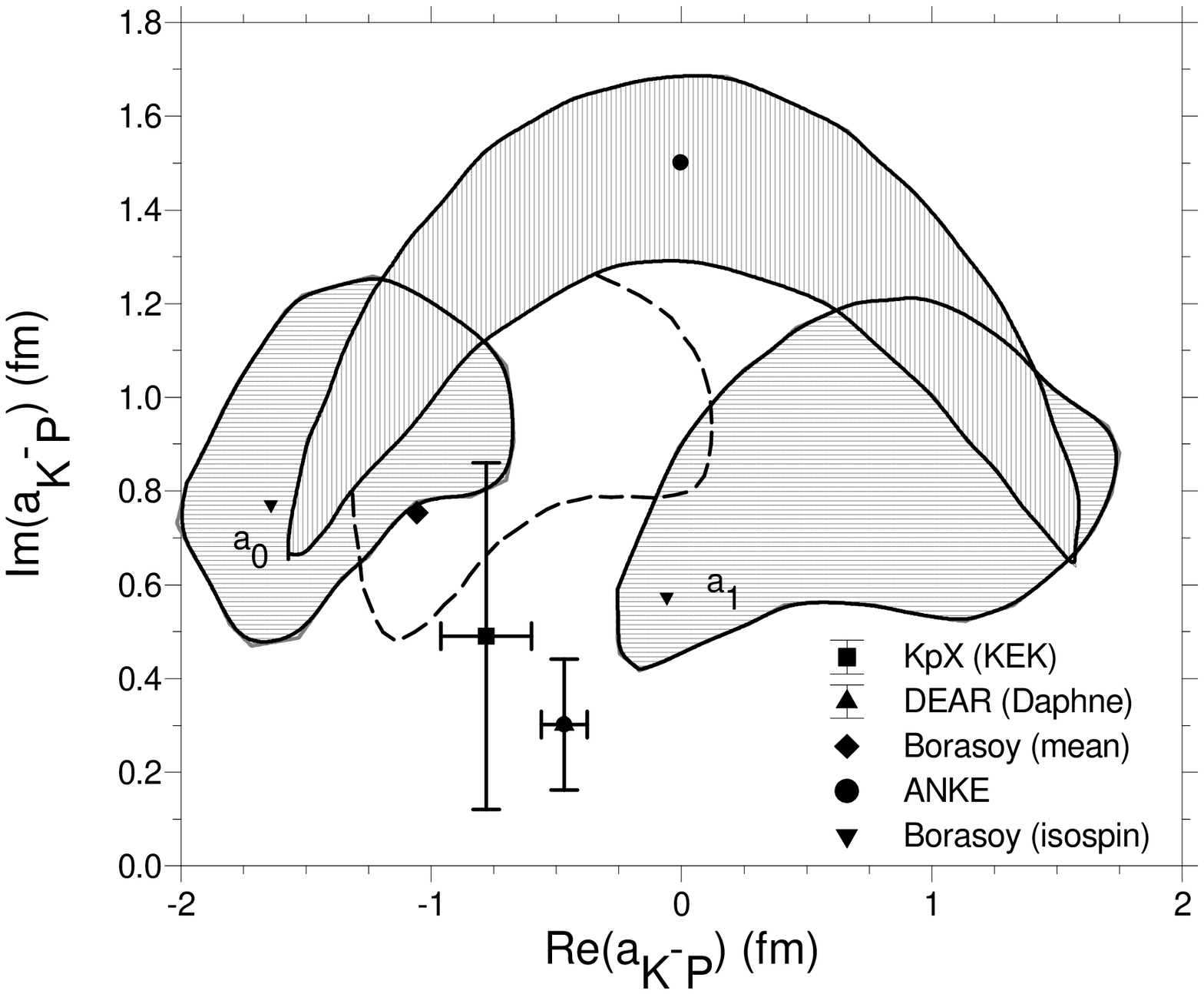}
\caption{Left panel: Total cross sections for kaon-pair production as a function of the
excess energy $\epsilon$ measured at COSY. Black: $pp\to ppK^+K^-$
\protect{\cite{Wol98,Que01,Win06,Har06,Mae08}}. Green: $pp\to dK^+K^0$
\protect{\cite{Kle03,Dzy06}}.
Red: $pn\to dK^+K^-$
\protect{\cite{Mae06,Mae09}}. Pink: $pd\to {^3\text{He}}K^+K^-$
\protect{\cite{Bel07}}. Blue: $dd\to {^4\text{He}} K^+K^-$
\protect{\cite{Yua09}}.
The high energy $pp\to ppK^+K^-$ result (open circle) has been measured
at  SATURNE \protect{\cite{Bal01}}. Right panel: The real and imaginary
part of the ${K^-p}$ scattering length $a_{K^-p}$. The \fsi result from
ANKE \cite{Mae08} is shown as dot and the one-$\sigma$ level
uncertainty as contour plot (vertical shaded area). The chiral (SU3)
calculation from Ref. \cite{Bor06} is shown as full diamond together
with its uncertainty contour (dashed curve).  Similarly the pure isospin
results are presented as open squares and contours (horizontal shaded area).
The scattering length from elastic  $K^-$ scattering on protons
are from the KEK experiment KpX \cite{Ito98} and from the Daphne experiment
DEAR \cite{Gua04} are also shown. } \label{Fig:Kaon_pair}
\label{Fig:Kaon_pair}
\end{center}
\end{figure}
They are from Refs. \cite{Wol98 ,Que01 ,Win06 ,Har06 ,Mae08, Kle03,
Dzy06, Mae06, Mae09, Bel07, Yua09, Bal01}. On a first view the cross
sections seem to follow an universal curve except for the $dd\to
K^+K^-\alpha$ reaction, which is completely off the other data. If
we ignore this point as well as those for the $pp\to K^+K^0d$
reaction, which are slightly above the other data points, one can
parameterize the cross section as $\sigma_{tot}=0.00137\epsilon^{2.376}$ nb with $\epsilon$ measured in MeV. The reason for this behavior are final state
interactions (\emph{FSI}) between various particles. For the $pp\to
K^+K^-pp$ reaction these are $pp$, $K^-p$ and $K^+K^-$ \emph{FSI}. The
deduced scattering lengths are $ a_{K^ -  p}  = (0 + 1.5i)$ fm (Ref.
\cite{Mae08}) and $ a_{K^ -  K^ +  }  = \left[ {0.5_{ - 0.5}^4  + (3
\pm 3)i)} \right]$ fm (Ref. \cite{Sil09}). A possible \emph{FSI}   in $K^+
p$ was found to be of no importance. While the phase in the complex
scattering length $ a_{K^ -  p}$ is quite uncertain, $| a_{K^ -
p}|$ is much better defined (see Fig. \ref{Fig:Kaon_pair} and Ref. \cite{Mae08}). The $K^-p$ interaction is quite complex
because of the channel couplings to  $\Sigma\pi$  and also because
there are two isospins $I = 0$ and $I = 1$. In a recent study
\cite{Bor06}, within a chiral $SU(3)$ unitary approach,
$a_0=(-1.64+0.75i)$ fm and $a_1=(-0.06+0.57i)$ fm were obtained.
Here the sign convention is that negative scattering length
corresponds to repulsion.  These values are shown, together with
their variances in Fig. \ref{Fig:Kaon_pair}. Also the experimental
result is shown.  However, it is unclear in which isospin channel
the $K^-p$ system is produced and whether isospin is conserved.
However, the isospin mean value and its uncertainty has almost no
overlap with the \emph{FSI} result. The same is true for scattering length values from elastic $K^-p$ scattering. These results from KEK
\cite{Ito98} and Daphne \cite{Gua04} are also shown (see Fig.
\ref{Fig:Kaon_pair}). In $pd\to K^+K^-{^3\text{He}}$ reaction no evidence is found for $K^\pm{^3\text{He}}$
\emph{FSI}. One reason might be that the MOMO experiment \cite{Bel07} did not distinguish between the two kaons leading to an average over the two possible final states.

\section{$\phi$-Meson Production}\label{sec:Phi-Meson}
\begin{figure}[t]
\includegraphics[width=0.465\textwidth]{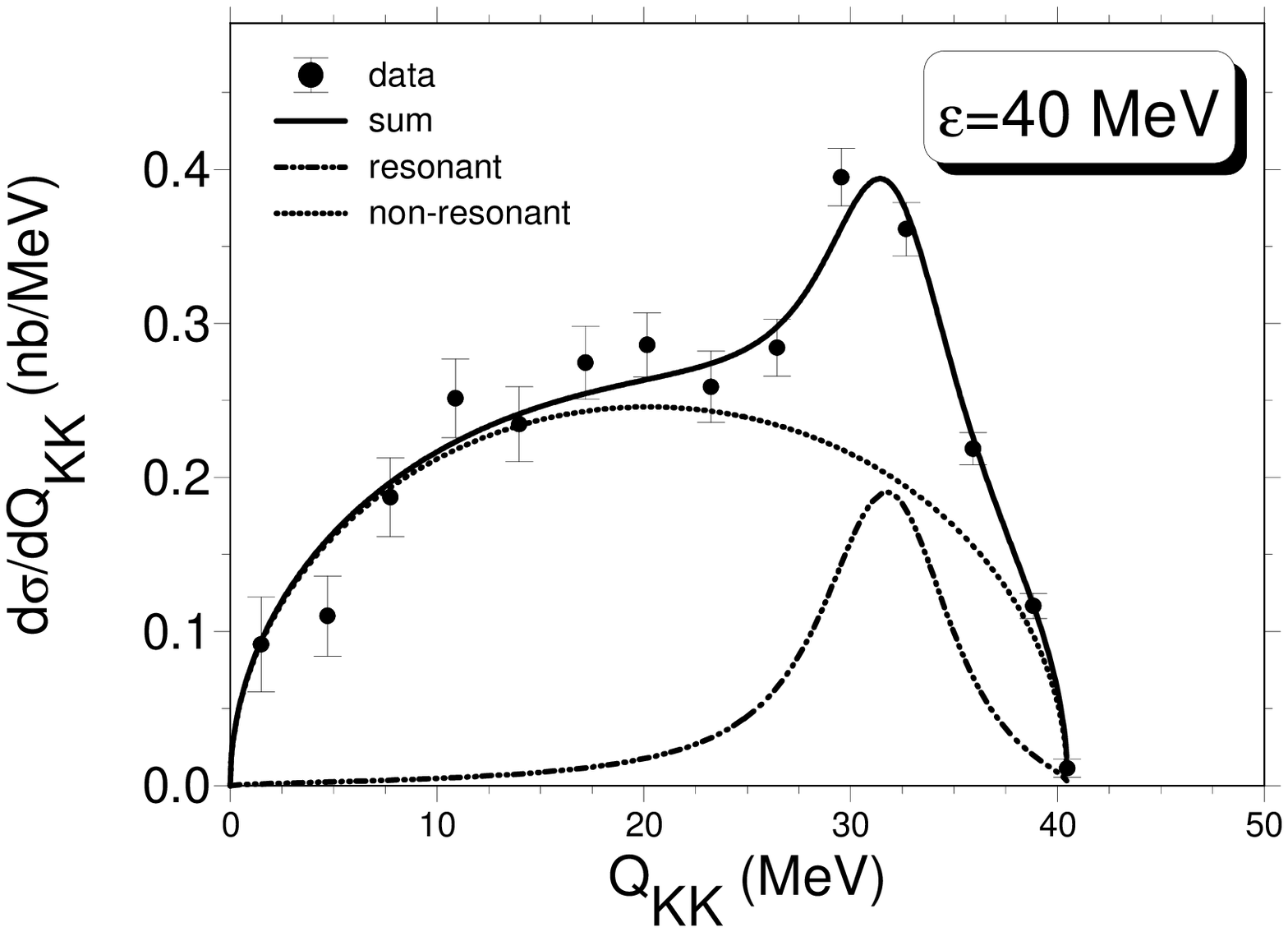}
\includegraphics[height=0.335\textwidth]{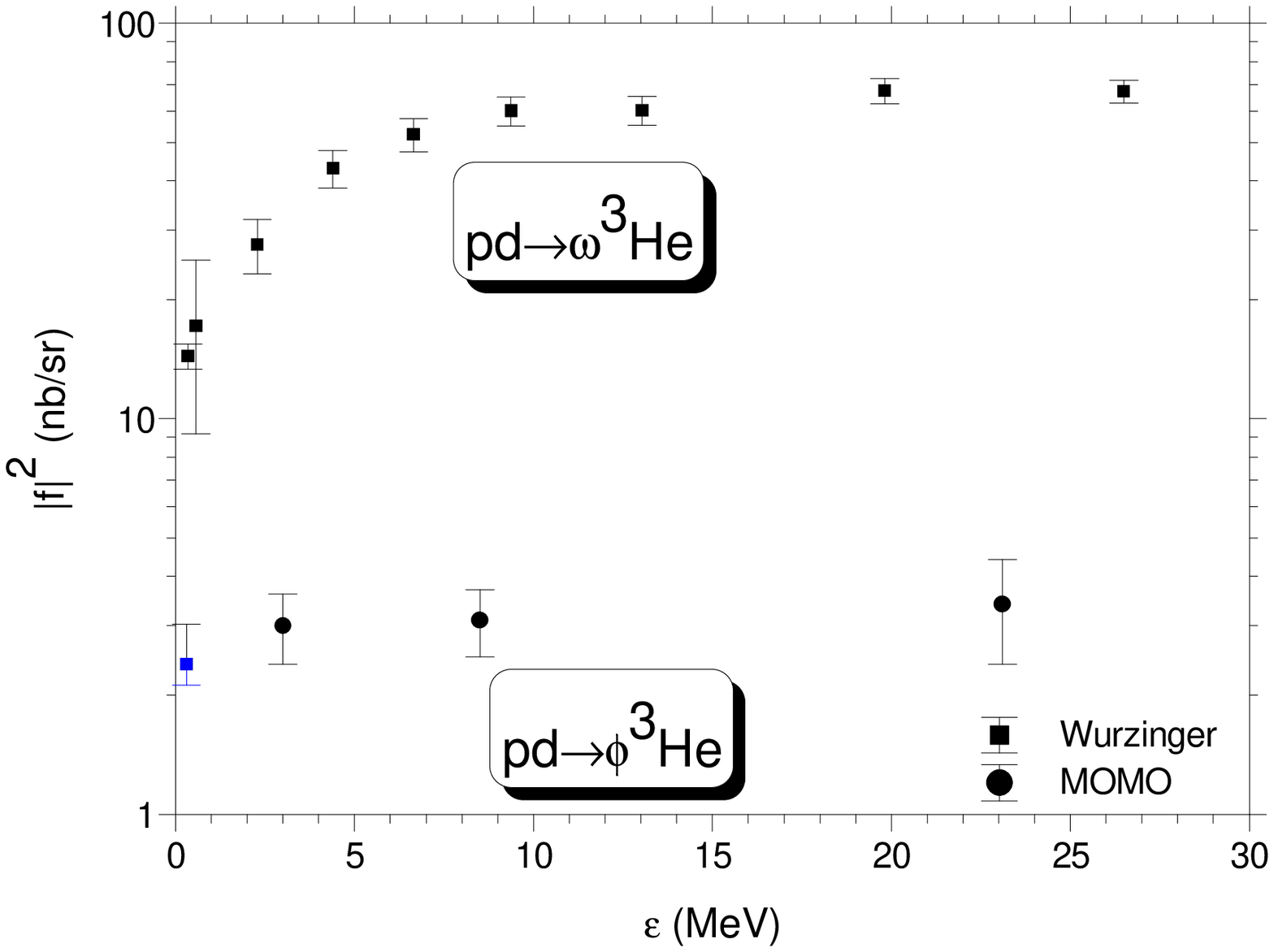}
\caption{Left panel: The differential cross section for the reaction $pd\to K^+K^-{^3\text{He}}$ for an excess energy of 40 MeV. The different curves are explained in the figure. Right panel: Excitation functions of $|f|^2$ for the two indicated reactions. The data labelled Wurzinger are from Ref. \cite{Wur96} and those labelled MOMO from Ref. \cite{Bel07} }
\label{MOMO}
\end{figure}

In addition to the non-resonant two-kaon production there is also a
resonant production possible: $pp\to \phi pp$ followed by the decay
$\phi\to K^+K^-$. The two processes can be distinguished on an
invariant mass plot of the two kaons. In Fig. \ref{MOMO} we show the
cross section as function of the energy between the two kaons for
the reaction $pd\to K^+K^-{^3\text{He}}$ at an energy 40 MeV above threshold.
The figure shows in addition to the data \cite{Bel07} the
non-resonant fraction as normalized phase space and the resonant
fraction by a Gaussian smeared Breit-Wigner distribution as well as
the sum of both. It is remarkable that in the case of $pp$ induced
reaction always $\sigma_{res}>\sigma_{nr}$ holds \cite{Mae08}. In the $pd$
induced reaction the opposite is true: $\sigma_{res}<\sigma_{nr}$.
There is one more distinct difference between both reactions. The
data were transformed into the Gottfried-Jackson frame \cite{GJ64}. In this
frame, which connects the entrance and exit channels, the total momentum of the $K^+K^-$ system is zero,
which means that it is the $\phi$ rest frame. Since the $\phi$ is a
vector meson, the distribution in the relative momentum of the kaons
from its decay is sensitive to its polarization with respect to some
quantization axis. For the $pp\to\phi pp$ reaction as well as for
the $pn\to\phi d$ reaction \cite{Mae06} it is found that $m=\pm 1$
and hence pseudoscalar meson exchange in the production graph \cite{Jackson64}. On the contrary in $pd\to \phi{^3\text{He}}$ the magnetic quantum number is
$m=0$ and hence vector meson exchange occurred. Why is the
production process so different? The  $pd\to \phi{^3\text{He}}$ reaction
measured was well below the $pp\to\phi pp$ production threshold.
Therefore, both nucleons in the deuteron had to coherently
participate in the reaction. It is surprising that in the  $pd\to
\omega{^3\text{He}}$ reaction no polarization was found \cite{Sch08}.

Another difference between these two reactions is the size of the cross section. From OZI rule a smaller $\phi$ production cross section is expected \cite{Lip76}. In Fig. \ref{MOMO} also the matrix elements squared $|f|^2=d\sigma/d\Omega /PS_2$ with $PS_2$ the two body phase space for the two reactions $pd\to{^3\text{He}}\phi,\omega$ are compared with each other. However, $\phi$ production is larger than predicted by the OZI rule: $R_{\phi/\omega}=8\times R_{OZI}$ for $pp$ reactions and only $R_{\phi/\omega}=20\times R_{OZI}$ for the $pd$ reaction with $R_{OZI}=4.2\times 10^{-3}$.

\section{Hyperon Production}\label{sec:Hyperon-Production}
\begin{figure}
\centering
\includegraphics[width=0.45\textwidth]{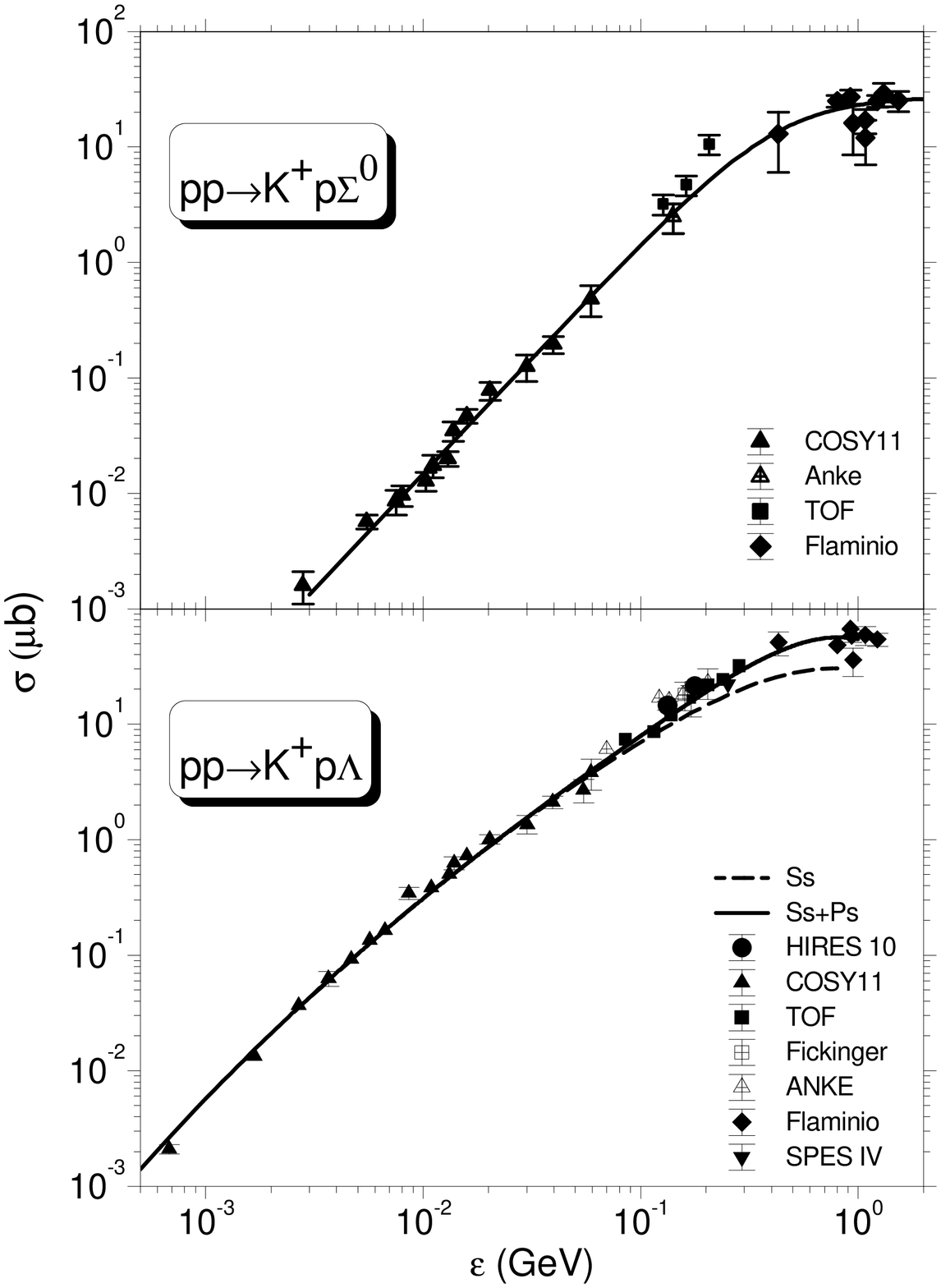}
\includegraphics[width=0.457\textwidth]{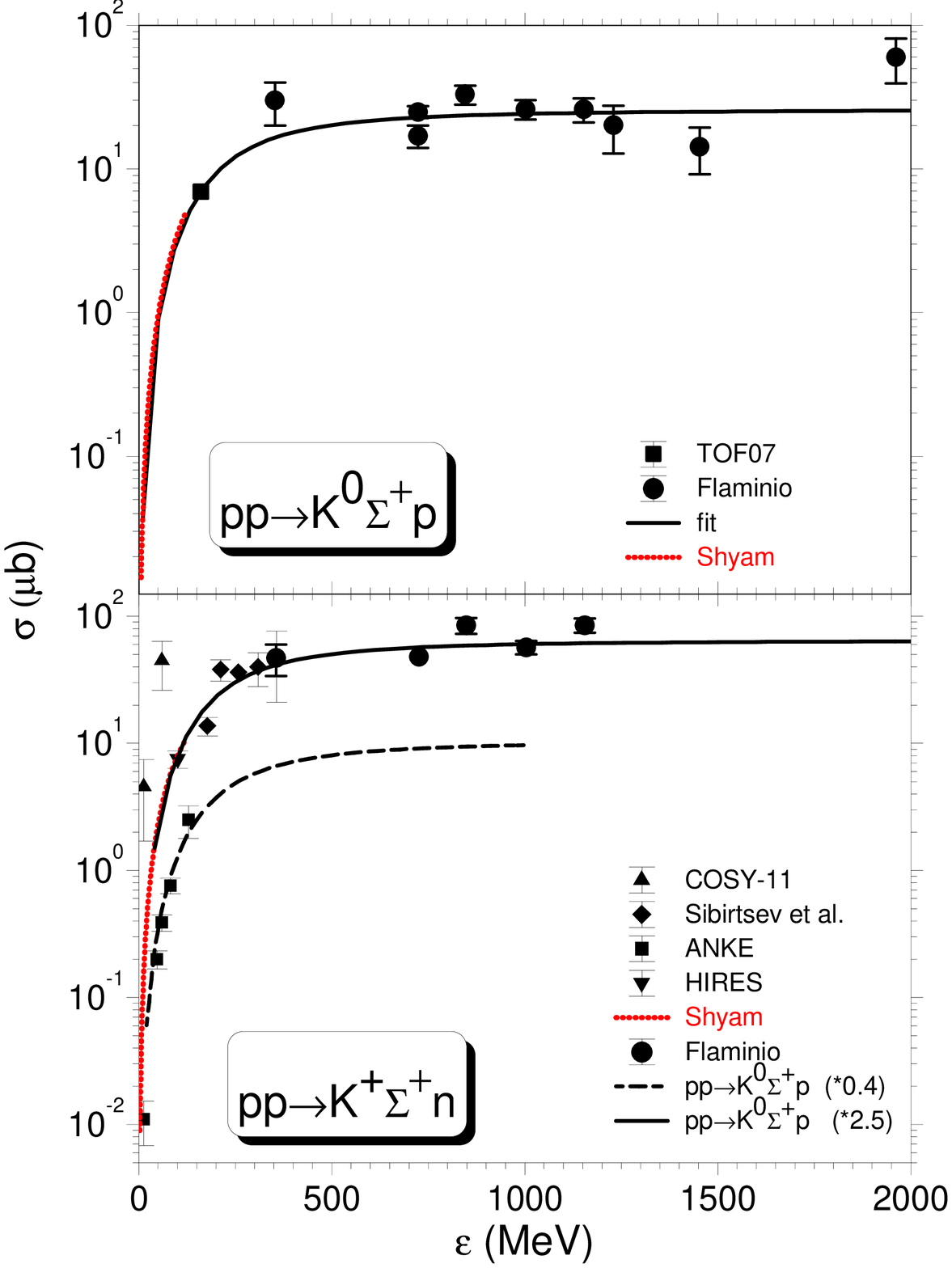}
\caption{Left panel: Excitation functions for the reactions $pp\to K^+Y^0p$ with $Y^0=\Lambda$ (lower frame) and $\Sigma^0$ (upper frame). The data obtained at COSY are from HIRES \cite{Bud10}, \cite{Bud10a}, COSY11 \cite{COSY11}, TOF \cite{TOF}, and ANKE \cite{Val07}, earlier data from  \cite{Fic62, Siebert94, Flaminio}. For curves see text. Right panel: Same as left panel but for $\Sigma^+$ production. The near threshold point (upper right frame) is from TOF \cite{TOF07}. The dotted curves are from Ref. \cite{Shyam06}. The total cross sections for the $pp\to K^+\Sigma^+n$ reaction are from Refs. \cite{Bud10a, Val07, Flaminio, Cosy11a, Sib06}. }
\label{Fig:KY0p}
\end{figure}

The production of hyperons is of special interest because of the
production mechanism as well as their interactions. In a recent high
resolution experiment the $pp\to K^+\Lambda p$ reaction was studied
close to the threshold \cite{Bud10}. It is this area which is most
sensitive to final state interactions. The analysis yielded almost no
spin triplet strength for this reaction. The effective range
parameters extracted were $a_s=-2.43^{+0.16}_{-0.25}$ fm and
$r_s=2.21^{0.16}_{-0.36}$ fm. The total cross sections for two beam energies were determined \cite{Bud10, Bud10a} and are shown in Fig. \ref{Fig:KY0p} together with other data from \cite{COSY11, TOF, Fic62, Siebert94, Val07} and \cite{Flaminio}. The data can be described in terms of only three transitions: ${^3P_0}\to{ ^1S_0\,s_0}$, ${^1S_0}\to{^3P_0\,s_0}$ and ${^3P_{0,1,2}}\to {^3P_{0,1,2}\,p_{0,1,2}}$ (see Refs. \cite{Bud10a} and \cite{GEM02} for details). The usual spectroscopic notation $^{2S + 1} L_j \,\,l_J$ is applied with $S$, $L$ and $j$ denoting the spin, angular momentum and
total angular momentum in the final two-proton system,
respectively. $J$ is the total angular momentum and $l$ is the angular momentum between the meson and the two-proton system. Only two final states of these namely $Ss$ and $Ps$ were found necessary to account for the data. In the other reactions with $\Sigma$ production no \emph{FSI}   is observed. Therefore, the cross sections can be accounted for by a function, which varies like phase space in the threshold region and approaches a constant for higher energies: $$\sigma(\epsilon)=1/(a+b/\epsilon^2).$$ It is remarkable that almost all data - except the old data at high energies - are from COSY experiments. For the production of $\Sigma^+$ hyperons the data situation is less favorable than for the $Y^0$ cases. There is only one point from TOF close to threshold for $pp\to K^0\Sigma^+p$ although with extremely small error bar \cite{TOF07}.
For the reaction $pp\to K^+\Sigma^+n$ there are much more data from Refs. \cite{Bud10a, Cosy11a, Val07}. However, there is serious disagreement between these data. Shyam \cite{Shyam06} predicted in an effective Lagrangian model the near threshold cross sections for these two channels. His prediction is undistinguishable from the fit in the case of $pp\to K^0\Sigma^+p$. We further noticed that the ratio  between the two channels in his predictions is constant 2.5. The fit curve for $pp\to K^0\Sigma^+p$ times 2.5 is shown in the lower frame against the $pp\to K^+\Sigma^+n$ cross section data. It is below the data from Ref. \cite{Cosy11a} but higher than the data from \cite{Val07}. However, it meets the HIRES result and the higher energy data.  A factor of 1/2.5 would be necessary to reproduce the ANKE data \cite{Val07}, but then the high energy data were not met.

\begin{figure}[t]
\begin{center}
\includegraphics[width=0.47\textwidth]{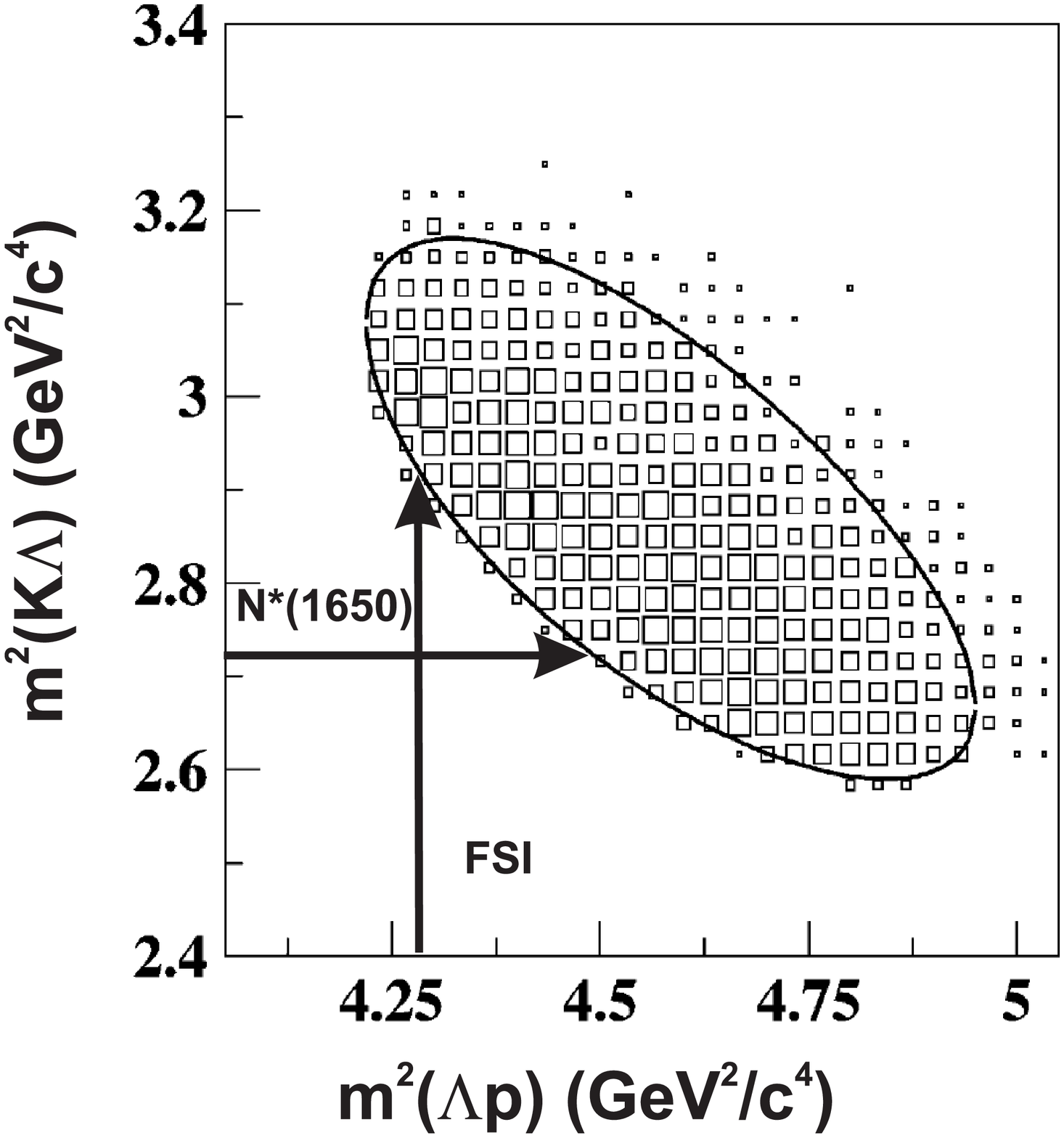}
\hspace{3mm}
\includegraphics[width=0.47\textwidth]{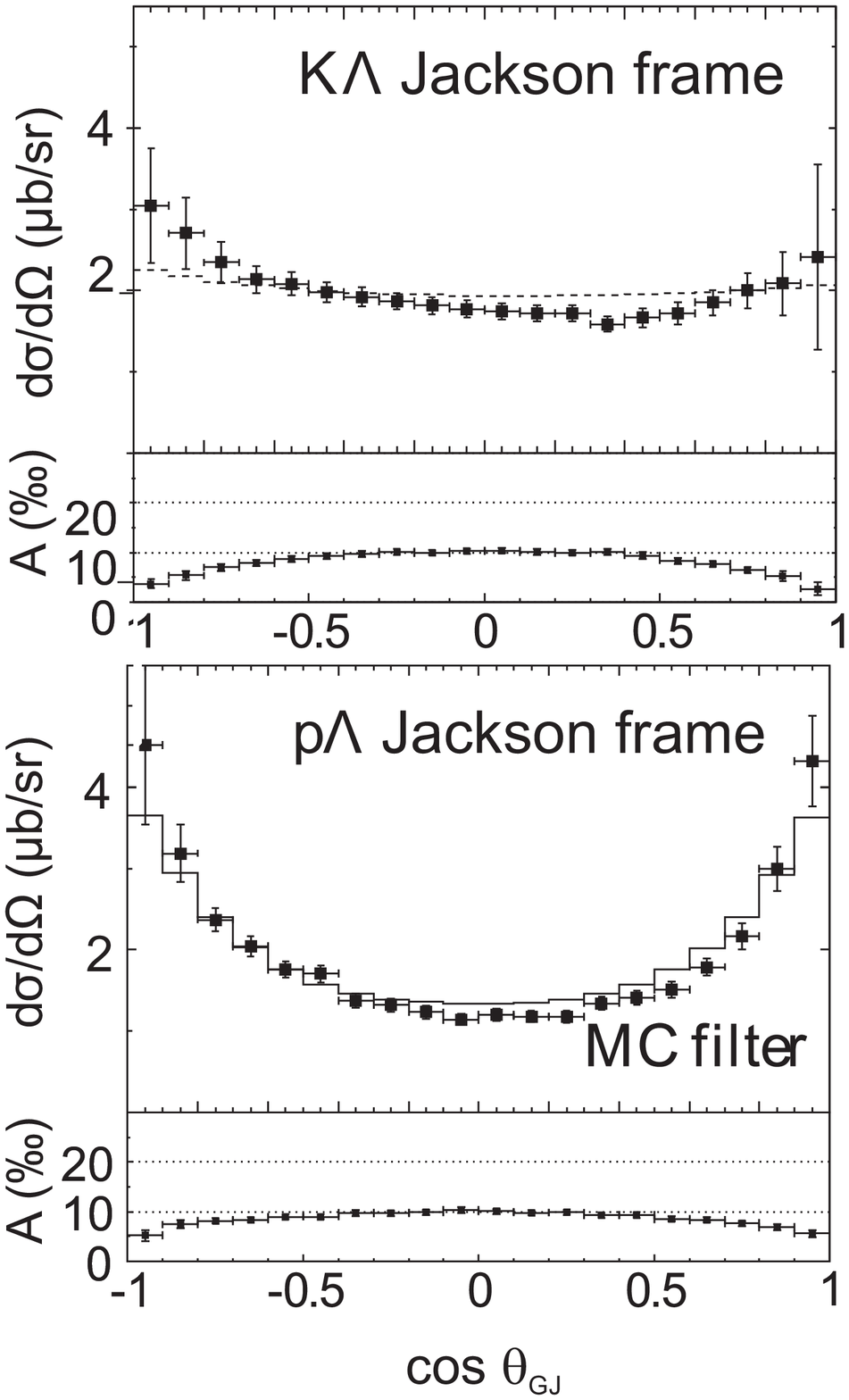}
\caption{Left panel: Dalitz plot of the reaction $pp\to K^+\Lambda p$ at 2.85 GeV/c from Ref. \cite{TOF}. The arrow on the $K^+\Lambda$ axis indicates the position ofthe $N^*$ resonance at $m(S_{11})=1.65\pm 0.15$ GeV/c$^2$. The arrow on the $\Lambda p$ axis indicates an enhancement due to \emph{FSI}. Right panel: Differential cross sections from $pp\to K^+\Lambda p$ at 2.55 GeV/c from Ref. \cite{TOF} transformed into the corresponding Gottfried-Jackson frame. The solid histogram in the $\Lambda p$ case is a Legendre polynomial fit. This is put into a Monte Carlo calculation yielding the dashed curve for the $K^+\Lambda$ case und limited angular momentum (see text). The frames below show always the differential acceptances.  }
\label{Fig:Dalitz}
\end{center}
\end{figure}
An interesting question to be answered by experiments is to what extent nucleon resonances contribute to hyperon production. In Fig. \ref{Fig:Dalitz} the Dalitz plot of the reaction $pp\to K^+\Lambda p$ at a beam momentum of 2.85 GeV/c is shown. The data are from TOF \cite{TOF}. An enhancement in the cross section is visible for the mass $m(K^+\Lambda)\approx 1.65$ GeV/c$^2$. The width of the resonance $N^*(1650)$ is $\Gamma=0.15$ GeV/c$^2$. This means that the resonance covers the full allowed kinematical range. Indeed, this resonance was found to dominate the near threshold cross section (see Refs. \cite{TOF, Shyam06}). It is interesting to note that the quantum numbers of this $N^*$ resonance are identical to those of the $N^*(1535)$: negative parity, spin and isospin 1/2. It plays the same role in kaon production as the $N^*(1535)$ in $\eta$ production.

Another way of analysis is the transformation of the data into a Gottfried-Jackson frame. Here we will concentrate on two possibilities. The first one is $\vec{p}_\Lambda=-\vec{p}_K$ which is the decay if the $N^*$ in its rest frame. The Gottfried-Jackson angle $\theta_{GJ}$ is then the angle between the average of beam and target proton direction and the $N^*$ direction. The angular dependence of the cross section is shown in the upper frame of the right panel in Fig. \ref{Fig:Dalitz}. An expansion in Legendre polynomials up to second order are sufficient to account for the data. Thus $l\le 1$ and only $S_{11}$, $P_{11}$, and $P_{13}$ resonances can contribute. In the case $\vec{p}_\Lambda=-\vec{p}_p$, shown in the frame below, it is found that only $l=0$ and $l=1$ contribute in consistency with the finding from the analysis of the excitation function in Fig. \ref{Fig:KY0p}. It should be stressed that these findings cannot be obtained from a Dalitz plot.

\section{Search for Resonances in the $\Lambda p$ Channel}\label{sec:Search-Resonan-Lambda-p-Channel}
%
%%%%%%%%%%%%%%%%%%%%%%%%%%%%%%%%%%%%%%%%%%%%%%%%%%%%%%%%%%%%%%%%%%%%%
%
\begin{figure}[h!!]
\begin{center}
\includegraphics[width=0.4\textwidth]{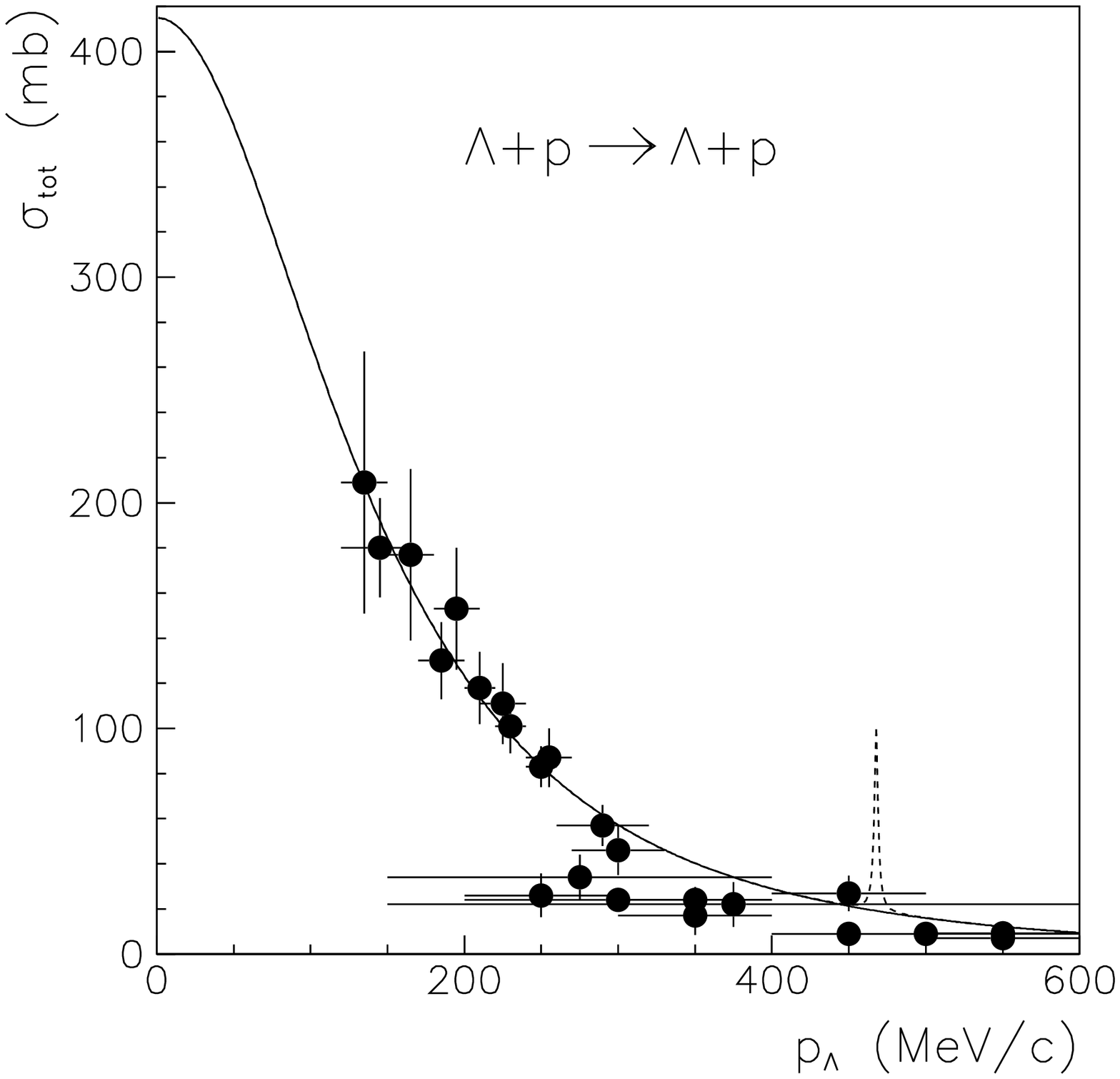}
\includegraphics[width=0.4\textwidth]{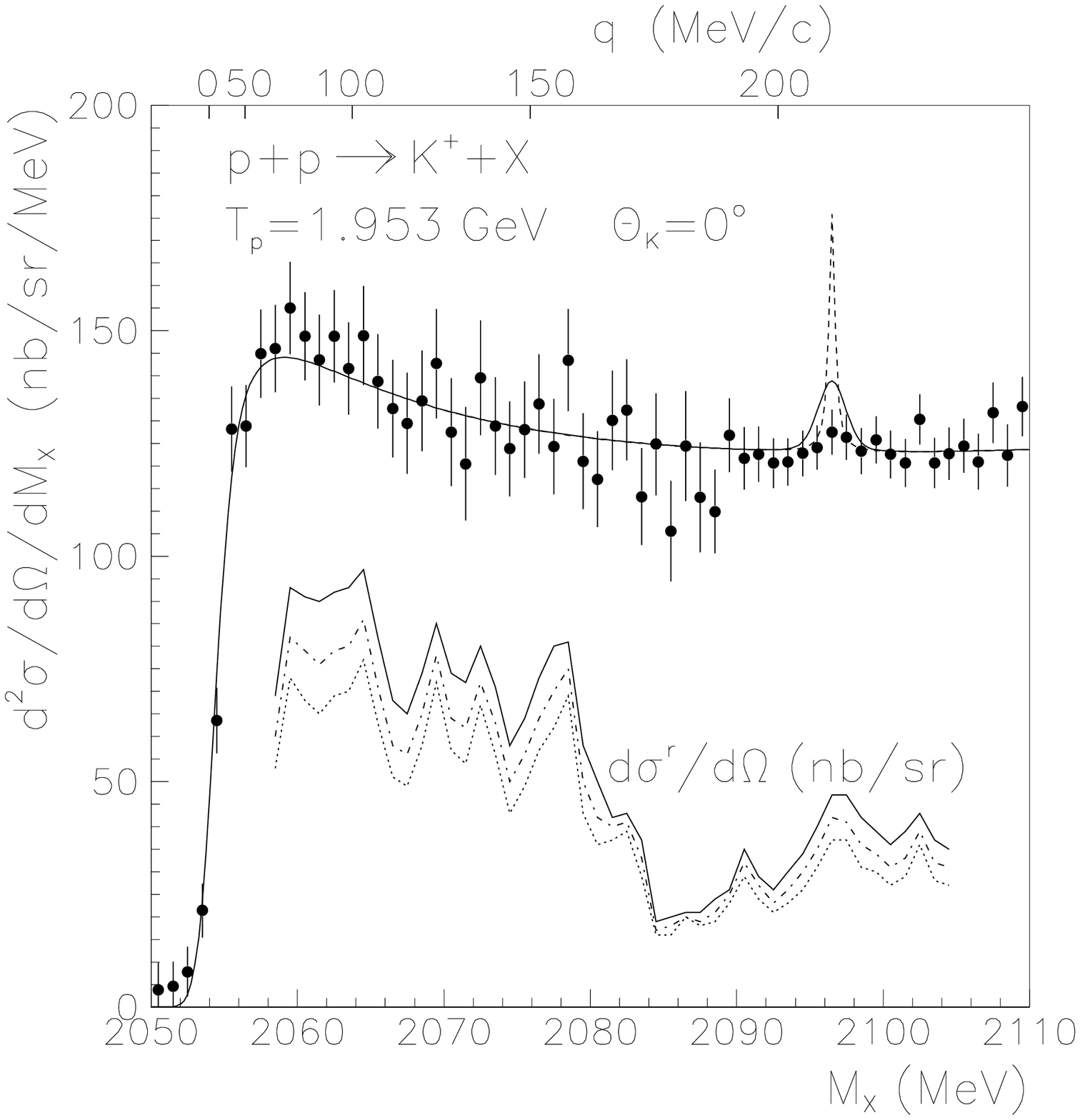}
\caption{Left panel: Total $\Lambda p \to \Lambda p$ cross section  vs. laboratory momentum $p_{\Lambda}$.
Solid line: effective range approximation of $\sigma_{tot}^{nr}$.
Dashed line: Simulation of a resonance excursion $\sigma_{tot}^r$
without folding with the effective resolution function
for a resonance in the $^1P_1$ channel with
$E_r=42.5$~MeV and $\Gamma=500$~keV. For data see \cite{Bud11}.
Right panel: Missing mass spectrum of the reaction $p+p \to K^+ + X$ with $X=(\Lambda p)$.
The upper axis indicates
the c.m. momentum $q$ of the $\Lambda p$ system. Solid line: \emph{FSI}   fit curve with
resonance excursion excluded by the $\chi^2$ test.
Dashed line: Same resonance excursion without folding with the effective resolution function.
Resonance parameters:
${\rm d}\sigma^r/{\rm d}\Omega_K=42$~nb/sr, $M_r=2096.5$~MeV, $\Gamma=500$~keV.
The lower curves represent upper limits of the production cross section
${\rm d}\sigma^r/{\rm d}\Omega_K$ (nb/sr)
are indicated for $\Gamma=100$ keV (dotted line), $\Gamma=500$ kev (dashed-dotted line) and
$\Gamma=1.0$ MeV (solid line).
}
\label{Fig:spectrum}
\end{center}
\end{figure}
Similar to a resonance in the $K^+\Lambda$ channel there might be a resonance in the $\Lambda p$ channel. Such a resonance would have strangeness $S=-1$. Aerts and Dover \cite{Aer85} predicted a spin singlet dibaryon $D_s$ between the $\Lambda p$ and the $\Sigma N$ thresholds. We searched for such a resonance in $\Lambda p$ elastic scattering as well as in \emph{FSI}   (Ref. \cite{Bud11}).  We assumed a resonance at 42.5 MeV having a width of 0.50 MeV. Such a resonance could, however, not been seen in the elastic cross sections (see left panel in Fig. \ref{Fig:spectrum}).  This resonance corresponds to a mass of $2.0965\pm 0.5000$ MeV/c$^2$ in \emph{FSI}. In a previous experiment \cite{Siebert94} some structure had been observed. This is not confirmed by the high resolution experiment \cite{Bud11}. A $\chi^2$ analysis leads to upper limits for the  cross sections for the $pp\to K^+D_s$ reaction (see right panel in Fig. \ref{Fig:spectrum}).

\section*{Acknowledgements}\label{sec:Acknowl}
Stimulating and enlightening discussions with colleagues from COSY are acknowledged. One of us (H. M.) is grateful to J. Niskanen for helpful discussions.
%
%%%%%%%%%%%%%%%%%%%%%%%%%%%%%%%%%%%%%%%%%%%%%%%%%%%%%%%%%%%%%%%%%%%%%
%%%%%%%%%%%%%%%%%%%%%%%%%%%%%%%%%%%%%%%%%%%%%%%%
% Begin References
%%%GATHER{c:\Program Files\MiKTeX 2.9\bibtex\bib\eigene\NN_Interaction.bib}
%

%%\bibliographystyle{elsart-num}
%%\bibliographystyle{My}
%%\bibliography{NN_Interaction}

%%%%%%%%%%%%%%%%%%%%%%%%%%%%%%%%%%%%%%%%%%%%%%%%%
% End References
%%%%%%%%%%%%%%%%%%%%%%%%%%%%%%%%%%%%%%%%%%%%%%%%%
\end{document}